\documentclass[12pt]{article}

\usepackage{amsmath}
\usepackage{graphicx}
\usepackage{amssymb}
\usepackage{amsfonts}
\usepackage{latexsym}
\usepackage{color,soul} 

\def\beq{\begin{eqnarray}}
\def\eeq{\end{eqnarray}}
\def\nn{\nonumber}                

\def\det{\,\mbox{det}\,}

\def\diag{\,\mbox{diag}\,}

\renewcommand{\vec}[1]{{\bf #1}}

\renewcommand{\diag}{\,\mbox{diag}\,}

\def\al{\alpha}
\def\be{\beta}

\def\ga{\gamma}
\def\de{\delta}

\def\la{\lambda}
\def\na{\nabla}
\def\pa{\partial}

\def\si{\sigma}
\def\om{\omega}
\def\ph{\varphi}

\def\Ga{\Gamma}

\def\La{\Lambda}

\def\Om{\Omega}

\begin{document}

\begin{center}

{\large\bf Covariant derivative of fermions and all that}
\vskip 4mm

Ilya L. Shapiro

\end{center}

\begin{center}
{\sl
Departamento de F\'{\i}sica, ICE, Universidade Federal de Juiz de Fora
\\
Campus Universit\'{a}rio - Juiz de Fora, 36036-900, MG, Brazil
\vskip 2mm

}

\vskip 2mm

{\sl E-mail:
\ \
ilyashapiro2003@ufjf.br}

\end{center}
\vskip 4mm

\begin{quotation}
\noindent
\textbf{Abstract.}
We present detailed pedagogical derivation of covariant derivative
of fermions and some related expressions, including commutator
of covariant derivatives and energy-momentum tensor of a free
Dirac field. On top of that, local conformal transformations for a
Dirac fermion in curved spacetime are considered and we obtain the
expression for the energy-momentum tensor on the cosmological
background.
\vskip 3mm

\vskip 2mm


\end{quotation}

\tableofcontents

\section{Introduction}

The formulation of classical fields on an arbitrary curved background
can be seen as an important element of general relativity. On the
other hand, this is one  of the first steps in constructing the quantum
field theory on a curved background
(see, e.g., \cite{DeWitt65,birdav,book,Rubakov,ParToms,OUP}).
The main difficulty is related to that, in general,
it is not possible to start from the usual consideration based on the
representations of the Lorentz group, because the general Riemann
space (one can call it metric background) has much less symmetries
compared to the Minkowski space.  One can expect to achieve a
systematic construction of the fields on de Sitter or anti-de Sitter
spaces (see, e.g., \cite{UM} and references therein)
but this is not what we actually need in both  classical and
semiclassical gravity, since the metric backgrounds of our interest
are much more general than these two examples.

The simplest possible solution for formulating classical fields on an
arbitrary metric background consists of the covariant generalization
of the flat space-time expressions. This means, one has to start from
the action of the field in the locally flat reference frame and
construct the covariant expression which reduce to the flat-space
action in this special frame. Such a procedure is sometimes called
covariantization, it is rather simple for scalars and vectors, and
essentially more involved for spinor fields. In what follows, we
consider the main steps in formulating such a covariant
generalization of the flat spacetime action for a Dirac fermion.
The action, in the flat case, is
\beq
S_f
&=&
i \int d^4x \,\,{\bar \psi}\big( \ga^a \pa_a - im \big) \psi\,,
\label{flat}
\eeq
where $\,a=0,1,2,3\,$ are Minkowski-space indices. Along with
this main problem, we shall present necessary details of the
tetrad formalism, calculate dynamical energy-momentum
tensor of spinor field and discuss the conformal properties of
this field. The last step will be derivation of the energy-momentum
tensor of spinor field on the cosmological background, in the case
when spinor depends only on time and not on the space coordinates.
All these calculations and considerations are pretty well-known and
we do not pretend at all to say a new word in this field. The purpose
of the work is mainly pedagogical, namely we intend to present the
detailed derivations which can be useful for the one who wants to
learn the subject.

The last note is that the original version of this manuscript was
supposed to be part of the book \cite{OUP}. However, in this book
another (albeit equivalent) scheme of constructing fermion fields in
curved space was chosen, based on the application of the group
theory methods.
The approach used in the mentioned book utilizes the fact that the
spinor connection, which will be defined below, is a connection for
a local Lorentz gauge group. This group is present also in curved
background and one can use it to derive both spinor connection
and the commutator of covariant derivatives. The interested reader
may find it useful to make a comparison between the two approaches.

The paper is organized as follows. Sec.~\ref{s2} discusses the
tetrad (also called vierbein) formalism, including the detailed
consideration of what means the covariant derivative of the tetrad.
In Sec.~\ref{s3} we describe the construction of the covariant
derivative of Dirac fermion.
Sec.~\ref{s4} is deriving the commutator of covariant derivatives
and its connection to the Riemann tensor. Sec.~\ref{s5} describes
local conformal transformations in the action of spinors in curved
spacetime. In this section, we consider the $n$-dimensional
spacetime, different from the rest of the paper dealing with $n=4$.
In Sec.~\ref{s6} we present a detailed derivation of the action of
Dirac fermion for a weak gravitational field, obtain the
energy-momentum tensor $T_{\mu\nu}$ for the Dirac field and
show the connection of its trace with the conformal symmetry.
Sec.~\ref{s7} shows the calculation of $T_{\mu\nu}$ on the
cosmological background and explains the problems of taking
the massless limit in the naive way. Finally, in Sec.~\ref{s8} we
draw our conclusions.
The notations include the signature
\ $\eta_{\al\be} = \diag (+\,-\,-\,-)$,
the definition of the Riemann tensor
\beq
R^\la_{\,\,\,\tau\al\be} \,=\,
\pa_\al\,\Ga^\la_{\,\,\tau\beta}
- \pa_\be\,\Ga^\la_{\,\,\tau\al}
+ \Ga^\la_{\,\,\ga\al}\,\Ga^\ga_{\,\,\tau\be}
- \Ga^\la_{\,\,\ga\be}\,\Ga^\ga_{\,\,\tau\al} ,
\label{curva}
\eeq
Ricci tensor $\,{R^\al}_{\mu\al\nu} = R_{\mu\nu}$, and its trace
$\,R=R_{\mu\nu} g^{\mu\nu}$, i.e., the Ricci scalar.
Finally, our notations for symmetrization can be easily understood
from the following two examples:
\beq
A_{(ij)}=\frac12(A_{ij}+A_{ji})
\qquad
\mbox{and}
\qquad
B_{(i|k|j)}=\frac12(B_{ikj}+B_{jki}).
\eeq

\section{Tetrad formalism and covariant derivatives}
\label{s2}

It is well-known that general relativity is a theory of the metric field.
At the same time, in many cases, it proves useful to use other variables
for describing the gravitational field. In particular, for defining the
covariant derivative of a fermion, one need an object called tetrad.
The German name {\it vierbein} is also frequently used. The names
indicate that the object is four-dimensional, but the formalism
described below can be easily generalized to the space of any
dimension and also does not depend on the signature of the metric.
Anyway, for the sake of definiteness we will refer to the
four-dimensional space-time with the $M_{1,3}$ signature.

Let us start by defining the tetrad formalism  on the $M_{1,3}$
Riemann space. Locally, at point $P$, one can introduce the flat
metric  $\eta_{ab}$. This means that the vector basis  includes
four orthonormal vectors $\vec{e}_a$, such that
$\vec{e}_a \cdot \vec{e}_b =\eta_{ab} $. Here $\vec{e}_a $ are
4-dimensional vectors  in the tangent space to the manifold  of
our interest at the point $P$. Furthermore, $X^a$ are local
coordinates on  $M_{1,3}$, which can be also seen as coordinates
in the tangent space in a close vicinity of the point $P$. With
respect to the general coordinates  $x^\mu$, we can write
$\vec{e}_a = e^\mu _a \vec{e}_\mu$, where $\vec{e}_\mu$ is a
corresponding local basis and $e^\mu _a $ are transition
coefficients from one basis to another.  Hence
\beq
\nonumber
&&
e^{\mu}_{a}
\,=\,
\frac{\pa x^\mu}{\pa X^a}\,,
\qquad
e^a_\nu = \frac{\pa X^a}{\pa x^\nu},
\quad
\mbox{and therefore}
\\
\label{1}
&&
e^{\mu} _a e^{a\nu}
\,=\,
e^\mu _a e^\nu _b \eta^{ab} = g ^{\mu \nu}\,,
\\
\nonumber
&&
e^{a} _{\mu} e_{a\nu}
\,=\,
g_{\mu \nu}\,,
\quad
e^a _\mu e^\mu _b
= \de^a _b\,,
\quad
e^a_\mu e^\al_{a} = \delta^\al _\mu\,.
\eeq
We assume that the Greek indices $\,\mu,\nu, \dots $ are raised and
lowered by the covariant metrics $ g_{\mu \nu}\,$ and $\,g^{\mu \nu}$,
while the indices $ a,b,c \dots $ are raised and lowered by $\,\eta _{ab}\,$
and $\,\eta ^{ab}$. The transition from one type or indices to another
is done using tetrad, since it is just a change of basis. Taking this into
account, in some cases, we admit objects with mixed indices, e.g.,
\beq
T^a_{\,\cdot\mu} &=& T^\nu_{\,\cdot\mu}\,e^a_\nu
\,=\,T^a_{\,\cdot b}\,e^b_\mu\,.
\label{mix}
\eeq

Formulas (\ref{1}) show that the descriptions in terms of the metric
and in terms of the tetrad are equivalent. The same concerns also the
invariant volume of integration. It is easy to show that
\beq
\det e^a_\mu = \sqrt{|g|}\,,
\qquad
g=\det g_{\mu\nu}\,,
\label{det}
\eeq
because
\beq
\label{2}
g = \det \big( e^a _\mu e^b _\nu \eta_{ab}\big)
&=&
\big(\det e^a _\mu\big)^2\cdot \det \eta_{ab} \, .
\eeq

Indeed, the local frame coordinates $\,X^a\,$ are not unique, since
even for the locally Minkowski metric one is allowed to make
transformations $\,{\vec e}^\prime _a = \La^b _{a\prime} e_b$,
such that\footnote{We use notations of the tensor textbook
\cite{Tensors}. In particular, prime over the index or over the vector
is the same thing, e.g., $\,V^\prime_\al = V_{\al^\prime}$.}
\beq
\label{3}
\vec{e}^{\prime} _a \cdot \vec{e}^{\prime} _c
&=&
\La^b _{a\prime} \La^d _{c\prime} \,\vec{e}_b \cdot \vec{e}_d
\,=\,
\eta_{bd} \,\La^b_{a\prime} \La^d_{c\prime} \,=\, \eta_{ac}\,.
\eeq
In the flat space-time case, such frame rotations with constant
$\,\La^b _{a\prime}\,$ form Lorentz group. The situation changes
in curved space-time manifolds, because  then $\La^a _{b'}$ depends
on the point $P$. For example, we  can consider an infinitesimal
transformation
\beq
\La ^a _{b^\prime}
&=&
\de^a _{b^\prime} + \Omega^a _{b\prime} (x)\,,
\eeq
which preserves the form of the tangent-space metric.  Then
(\ref{3}) gives
\beq
\label{4}
(\delta ^{a} _{b\prime} + \Omega^a _{b\prime})(\delta^d _{c\prime}
+ \Omega^{d}_{c\prime}) \,\eta_{ad} &=& \eta_{bc} \,.
\eeq
It is an easy exercise to show that  in this case,
$\,\Om_{ab}(x) \,=\, - \,\Om_{ba}(x)$.

 Let us consider the  covariant derivative in tetrad formalism
 and introduce the notion of spin connection. In the metric
 formalism, the covariant derivative of a vector  $\,V^{\mu}\,$ is
\beq
\label{5}
 \na_\nu V^\mu
 &=&
 \pa_\nu V^\mu + \Gamma^\mu_{\la\nu} V^\lambda\,,
\eeq
where the affine connection is given by the standard Christoffel
symbol expression,
\beq
\label{6}
\Ga^\la_{\mu\nu}
&=& \frac{1}{2}\, g^{\la\tau} \big(
\pa_\mu g_{\tau\nu} +\pa_\nu g_{\tau\mu} - \pa_\tau g_{\mu\nu}\big)\, . 		
\eeq
Our purpose is to construct a version of covariant derivative for
the objects such as tetrad, which have local Lorentz indices.
Obviously, this is a non-trivial task, because tetrad is not a tensor
and therefore its covariant derivative is not a conventional
object. Hence our constructions will necessary involve certain
amount of {\it ad hoc} assumptions. In what follows, we shall
see that there is a scheme of consistent consideration of the
problem.

It is easy to understand that (different from the flat space-time)
in the Lorentz frame $X^a$ the covariant derivative of the same
vector cannot be just equal to $\pa_a V^b$, because otherwise
we come to the contradiction. The details will become clear below,
let us just say that qualitatively the reason is that any shift from
the point $P$ means the change from one tangent space to
another and hence deriving the corresponding difference
between two values of the vector requires an additional
definition. Let us suppose that the desired covariant derivative
is a linear operator and satisfies Leibniz rule. The general
expression satisfying these conditions is
\beq
\label{7}
\na_a V^b
&=&
\pa_a V^b + {\tilde \om}^a _{\,\cdot\, ca} V^c\,,	
\eeq
where $\,{\tilde \om}^a _{\,\cdot\, ca}\,$ are some unknown
coefficients. Later on we will discuss their relation to the spin
connection. In the flat space-time $\,{\tilde \om}^a _{\,\cdot\, ca}=0$.

We request that the vector components satisfy
$\,V^\mu = e^\mu _a V^a\,$ and the tensor components
satisfy $\,\na_\nu V^\mu = e^a _\nu e^\mu _b \,\na_a V^b $.
Then, according to (\ref{5}) and  (\ref{7}), we have
\beq
 \na_\la V^\mu
 &=&
 \pa_\la V^\mu + \Gamma^\mu _{\tau \lambda} V^b
 \,=\, e^a _\la e^\mu _b\, \na_a V^b
 \,=\, e^a _\la e^\mu _b \,
 \big(\pa_a V^b + {\tilde \om}^b _{\,\cdot\, ca} V^c)
\label{cova}
 \\
&=&
 \de^\mu _\tau \pa_\la V^\tau
 + V^\tau e^a _\la e^\mu _b \pa_a e^b _\tau
 + e^\mu _b e^c _b e^c _\tau
 V^\tau {\tilde \om}^b _{\,\cdot\, c\la}
 \nonumber
 \\
&=&
 \pa _\la V^\mu + V^\tau (e^\mu _b \pa _\la e^b _\tau
 + e^\mu _b e^c _\tau {\tilde \om}^b _{\cdot\, c\la})\, .
 \nonumber
  \eeq
Therefore, we arrive at the equation for
$\,{\tilde \om}^b _{\cdot\, c\la}$,
\beq
\label{8}
\Ga^\mu _{\tau \la}
&=&
e^\mu _b \,\pa_\la e^b _\tau
+ e^\mu _d e^c _\tau\, {\tilde \om}^d _{\cdot \,c\la}\,.
\eeq
Multiplying the last equation by $\,e^a _\mu e^{\tau b}$,
we arrive at the following solution:
\beq
\label{9}
{\tilde \om}^{ab} _{\,\cdot \cdot \,\la}
&=&
e^a _\mu e^{\tau b} \Ga^\mu _{\tau \la}
-  e^{\tau b} \pa_\la e^a_\tau\,.
\eeq

One of the important features of the last expression is antisymmety
in $(a,b)$, that means $\,{\tilde \om}^{ab} _{\,\cdot \cdot \mu}
=-{\tilde \om}^{ba} _{\,\cdot \cdot \mu}$.
In order to see this, consider the sum
\beq
\label{10}
\nonumber
{\tilde \om}^{ab} _{\,\cdot \cdot \mu}
+ {\tilde \om}^{ba} _{\,\cdot \cdot \mu}
&=&  
e^a _\nu e^{\la b} \Ga^\nu _{\la \mu}
+ e^b _\nu e^{\la a} \Ga^\nu _{\la \mu}
- e^{\la b} \pa _\mu e^a _\la - e^{\la a} \pa _\mu e^b _\la
\\
\nonumber
&=&
\frac12\,(\pa _\la g_{\nu \mu} + \pa _\mu g_{\nu \la} - \pa _\nu
g_{\mu \la}) \cdot  (e^{a\nu} e^{\la b} + e^{a \la} e^{\nu b})
 - e^{b \la} \pa _\mu e^a _\la - e^{a \la} \pa _\mu e^b _\la
\\
\nonumber
&=&
e^{a \nu} e^{\la b} e^c _\nu \pa _\mu e_{c \la}
+ e^{a \nu} e^{\la b} e_{c \la} \pa _\mu e^c _\nu
- e^{\la b} \pa_\mu e^a _\la - e^{a \la} \pa _\mu e^b _\la
\, =\, 0\,.
\eeq

Another observation follows from Eqs. (\ref{5}) and (\ref{7}).
We can rewrite (\ref{5}) in the form
\beq
\label{11}
\na_\nu (V^a e^\mu _a)
&=&
\pa _\mu (V^a e^\mu _a)
+ \Ga^\mu _{\la \nu} \,V^a e^\la _a
\eeq
and construct the ``covariant derivative'' of the tetrad
$\,\na_\nu e^\mu_a\,$ in a way consistent with the Leibnitz
rule and with both (\ref{11}) and (7). Then
\beq
\na_\nu V^\mu
&=&
\pa _\nu V^\mu
+ \Ga^\mu _{\la \nu} V^\la
\,=\,
\na_\nu(V^b e^\mu _b)
\,=\,
e^a _\nu \nabla _a (V^b e^\mu _b)
\label{cov}
\\
&=&
e^a _\nu e^\mu _b \nabla _a V^b
+ e^a _\nu V^b \na_a e^\mu _b
\,=\, e^a _\nu e^\mu _b (\pa _a V^b
+ {\tilde \om}^b _{\cdot \,ca} V^c) + e^a _\nu V^b \na_a
e^\mu _b\,.
\nonumber
\eeq
At the same time, we have
\beq
\pa _\nu V^\mu
&=&
e^a _\nu \pa_a (V^b e^\mu _b)
\,=\,
e^a _\nu e^\mu _b \pa _a V^b + V^b e^a _\nu \pa _a e^\mu _b \,.
\label{part}
\eeq
By combining (\ref{cov}) and (\ref{part}) one can easily obtain
the relation
\beq
V^\la e^c _\la \,{\tilde \om}^b _{\,\cdot ca} e^a _\nu e^\mu _b
\,+\,
e^a _\nu e^b V^\la \na_a e^\mu_b
&=&
V^\la e^b _\la e^a _\nu \pa _a e^\mu _b
+ V^\la \Ga ^\mu _{\la \nu}\, .
\eeq
Since this relation should be true for all $\,V^\la$, we arrive at
\beq
e^a _\nu e^b _\la \nabla _a e^\mu _b
&=&
e^a _\nu e^b _\la \,\pa _a e^\mu _b
+ \Ga^\mu _{\la \nu}
- e^a _\nu e^c_\la e^\mu _b \,{\tilde \om}^b _{\cdot \,ca}\,.
\eeq
Multiplying this equation by $\,e^\nu _d e^\la _e\,$ one can get
(changing some indices)
\beq
\na _a e^\mu_b
&=&
\pa _a e^\mu _b + \Ga ^\mu _{\la \nu} e^\la _b e^\nu _a
- {\tilde \om}^{cb} _{\,\cdot \cdot \,\nu} e^\nu _a e^\mu _c\,.
\eeq
Since this is a tensor quantity, we can multiply it by
$e^a _\tau$ and obtain the final result
\beq
\na_\tau e^\mu _b
&=&
e^a _\tau \pa _a e^\mu _b
+ \Ga^\mu _{\la \tau} e^\la_b
- {\tilde \om}^c _{\,\cdot \,b \tau}\, e^\mu _c \, .
\label{covatetra}
\eeq
One can show that, also\footnote{This can be done by either
repeating the steps leading to (\ref{covatetra}) or by directly
using (\ref{covatetra}). The interested reader can compare these
two ways to make this calculation. },
\beq
\na _\tau e_{\mu a}
&=&
e^b _\tau \pa _b e_{\mu a}
- \Ga^\la _{\mu \tau} e^\la _a
- {\tilde \om}^{\,\,\,\,b} _{a\,\cdot \, \tau}\, e_{\mu b} \, .
\label{covatetra-1}
\eeq
Finally, by direct replacement of (\ref{9}) one can easily check
that both covariant derivatives vanish, $\,\na_\tau e^\mu _b=0$
and $\,\na _\tau e_{\mu a}=0$. This fact is nothing else but a direct
consequence of the metricity property of covariant derivative,
because
\beq
\na _\tau g_{\mu\nu} &=&
\na _\tau \big(e_\mu^a e_\nu^b \eta_{ab}\big)
\,=\,
2\eta_{ab}\, e_\mu^a \,\cdot\,\na _\tau  e_\nu^b  \,=\,0\,.
\label{metric}
\eeq

\section{Covariant derivative of Dirac fermion}
\label{s3}

Let us construct a covariant generalization of the action of a
spinor field (\ref{flat}). The desirable expression has the form
\beq
S_f &=&
i \int d^4x \sqrt{-g} \,
{\bar \psi}\big( \ga^\mu \na_\mu - im \big) \psi\,.
\label{Dirac}
\eeq
Since the part related to the integration volume element is
relatively simple, what has to be done is to generalize the
gamma-matrices and to construct the covariant derivative.

Let us use tetrad and define  the curved-space gamma-matrices
as $\,\ga^\mu = e^\mu _a \ga^a\,$. Now the indices of the
the new gamma-matrices are lowered and raised by means of
the covariant metrics $\,g_{\mu\nu}\,$ and  $\,g^{\mu\nu}$.
It is easy to see that the new gamma-matrices satisfy the
curved-space version of Clifford algebra,
\beq
\ga_\mu\ga_\nu + \ga_\nu\ga_\mu &=& g_{\mu\nu}\,.
\label{Cliff}
\eeq
The next observation is that since $\,\ga^\mu = e^\mu _a \ga^a\,$ and
$\,\ga^a$ are constant matrices (see Appendix for the details), the
vanishing covariant derivative
$\,\na_\al e^\mu_b=0$ means $\,\na_\al \ga^\mu=0$.

The construction of covariant derivative is a little bit more
involved. Let us define
\beq
\na_\mu \psi
 &=&
 \pa_\mu \psi
 + \frac{i}{2}\,
 \om^{\,\,\,ab}_{\mu\, \cdot \cdot}\,\si_{ab}\,\psi \,,
 \quad
 \mbox{where}
 \quad
 \si_{ab} = \frac{i}{2} \big(\ga_a \ga_b - \ga_b \ga_a\big)\,.
\label{co1}
\eeq
 One can regard (\ref{co1}) as a hypothesis which can be proved
 or disproved. $\,\om^{\,\,\,ab}_{\mu\, \cdot \cdot}\,$  are
 antisymmetric coefficients of the {\it spinor connection}, which
 have to be found from the requirement of covariance.
Taking a conjugate of (\ref{co1}), we arrive at
\beq
\na_\mu \bar{\psi}
&=&
\pa_\mu \bar{\psi} - \frac{i}{2}\,
\om^{\,\,\,ab}_{\mu\, \cdot \cdot} \,\bar{\psi} \,\si_{ab} \, .
\label{conj}
\eeq

Using (\ref{co1}) and (\ref{conj}), it is easy to verify an
identity
\beq
\na_\al \big( \bar{\psi}{\psi}\big)
=\pa_\al \big( \bar{\psi}{\psi}\big),
\eeq
which is a natural result for a scalar combination $\bar{\psi}{\psi}$.

In order to obtain the equation for spinor connection,
consider the relation for a composite vector
\beq
\na_\mu (\bar{\psi} \ga^\al \psi)
&=&
\pa_\mu (\bar{\psi} \ga^\al \psi)
+ \Ga^\al _{\nu \mu} \bar{\psi} \ga^\nu \psi \,.
\label{vec}
\eeq
A simple calculation using Leibniz rule yields
\beq
&&
\na_\mu \bar{\psi} \cdot \ga^\al \psi
\,+\,
\bar{\psi}(\na_\mu \ga^\al)\psi
+ \bar{\psi} \ga^\al \na_\mu \psi
\,=\, \pa_\mu \bar{\psi} \cdot \ga^\al \psi
+ \bar{\psi} \pa_\mu \ga^\al \psi
+ \bar{\psi} \ga^\al \pa_\mu \psi
\nonumber
\\
&&
+ \,\Ga^\al _{\nu \mu} \bar{\psi} \ga^\nu \psi
\,-\,\frac{i}{2} \om^{\,\,\,ab}_{\mu\, \cdot \cdot}
\,\bar{\psi}\, \si_{ab} \ga^\al\psi
+\frac{i}{2} \om^{\,\,\,ab}_{\mu\, \cdot \cdot}
\bar{\psi}\ga^\al \si_{ab} \psi
\,=\,
\bar{\psi} \ga^a (\pa_\mu e^\al _a) \psi
+ \bar{\psi} \Ga^\al _{\nu \mu} \ga^\nu\psi\, .
\qquad
\nonumber
\eeq
As this equality should be valid for any field $\psi$,
we arrive at
\beq
&&
-\, \frac14\,\om^{\,\,\,ab} _{\mu\,\cdot\cdot} \,
e^\al_c \, \big[\ga^c (\ga_a\ga_b - \ga_b\ga_a)
- (\ga_a\ga_b - \ga_b\ga_a)\ga^c\big]
\,=\,
\ga^c \big(e^\nu _c \Ga^\al _{\mu\nu}
+ \pa_\mu e^\al _c\big) \,.
\label{eq1}
\eeq
By means of the well-known relations
\beq
&&
\ga^c \ga_a \ga_b
\,=\, 2 \de^c _a \ga_b - \ga_a \ga^c \ga_b\,,
\nonumber
\\
&&
2\de^c _a \ga_b - 2\de^c _b \ga_a + \ga_a \ga_b \ga^c - \ga^c \ga_b \ga_a
\,=\, - 2\de^c _b \ga_a + 2\ga_b \de^c _a - \ga_b \ga_a \ga^c
\nonumber
\eeq
we get
\beq
\om^{\,\,\,ab}_{\mu\, \cdot \cdot}
\,\big(e^\al_b \ga_a - e^\al_a \ga_b \big)
\,=\, \ga^c
\big(e^\tau _c \Ga^\al _{\tau \mu} + \pa_\mu e^\al _c\big)\,.
\eeq
The last equation can be easily solved owing to the antisymmetry
of $\,\om^{\,\,\,ab}_{\mu\, \cdot \cdot}$, the result is
\beq
\om^{\,\,\,ab}_{\mu\, \cdot \cdot}
&=&
-\,\frac12\,{\tilde \om}^{ab}_{\,\cdot \cdot\, \mu}
\,=\,
\frac{1}{2}\,(e^{b}_\tau e^{\la a}
\Ga^\tau_{\la \mu} - e^{\la a}\pa_{\mu}e^{b}_{\la})
\nonumber
\\
&=&
\frac14\,(e^{b}_\tau e^{\la a} - e^{a}_\tau e^{\la b})
\Ga^\tau_{\la \mu}
\,+\,
\frac{1}{4}(e^{\la b}\pa_{\mu}e_{\la}^{a}
- e^{\la a}\pa_{\mu}e^{b}_{\la})\,.
\label{spi}
\eeq
Here we established the relation with Eq.~(\ref{9}) and
presented both compact and explicitly antisymmetric forms
of the spinor connection.
\	
Finally, the expression for the covariant derivative is
(\ref{co1}) with the spinor connection (\ref{spi}).
\
At this point, we can say that the construction of the action
(\ref{Dirac}) is complete. In the next sections, we will derive
and discuss some additional relevant formulas and features.

\section{Commutator of covariant derivatives}
\label{s4}

The next necessary step is to derive the commutator of two
covariant derivatives acting on a Dirac spinor. Consider
\beq
\nonumber
\na_{\nu}\na_{\mu}\Psi
&=&
\na_\nu \big(
\pa_{\mu}\Psi + \frac{i}{2}\,\om^{\,\,\,ab}_{\mu\,\cdot\cdot} \si_{ab}\Psi\big)
\\
\nonumber
&=&
\pa_{\nu}(\na_{\mu}\Psi) + \frac{i}{2}\,\om^{\,\,\,ab}_{\nu\,\cdot\cdot}
\,\si_{ab}\,\na_{\mu}\Psi
- \Ga^{\la}_{\mu\nu}\na_{\la} \Psi
\\
\nonumber
&=&
\pa_{\nu}\pa_{\mu}\Psi
+ \frac{i}{2}\,\pa_{\nu}\om^{\,\,\,ab}_{\mu\,\cdot\cdot}
\,\si_{ab}\Psi
+ \frac{i}{2}\,\om^{\,\,\,ab}_{\mu\,\cdot\cdot}
\,\si_{ab}\,\pa_{\nu}\Psi
+ \frac{i}{2}\,\om^{ab}_{\nu}\,\si_{ab}\, \pa_\mu \Psi
\\
&-&
\frac{1}{4}\,\om^{\,\,\,ab}_{\nu\,\cdot\cdot}
\si_{ab}\,\om^{\,\,\,cd}_{\mu \,\cdot \cdot}\si_{cd}\Psi
-\Ga^{\la}_{\mu\nu}\pa_{\la}\Psi
- \frac{i}{2}\,\Ga^{\la}_{\mu\nu}\,
\om^{\,\,\,ab}_{\la \,\cdot \cdot}\,\si_{ab}\Psi\,.
\label{two}
\eeq
By using this expression and the relation
\beq
\ga_a\ga_b\ga_c\ga_d
&=&
 2\eta_{bc}\ga_{a}\ga_{d} - 2\eta_{ac}\ga_{b}\ga_{d}
 + 2\ga_{c}\ga_{a}\eta_{bd} - 2\ga_{c}\ga_{b}\eta_{ad}
 + \ga_{c}\ga_{d}\ga_{a}\ga_{b},
\label{gammas}
\eeq
it is not difficult to obtain the commutator
\beq
 [\na_{\mu},\na_{\nu}] \psi
 &=&
\na_\mu \na_\nu\psi - \na_\mu \na_\nu\psi
\,=\,-\,\frac14\,
R_{\mu\nu\,\cdot \cdot}^{\,\,\,\,\,\,\,ab}\,\ga_a\ga_b\, \psi\,,
\label{comm}
\eeq
where
\beq
R_{\mu\nu\,\cdot \cdot}^{\,\,\,\,\,\,\,ab}
&=&
\pa_\mu \om^{\,\,\,ab}_{\nu \,\cdot \cdot}
- \pa_\nu \om^{\,\,\,ab}_{\mu \,\cdot \cdot}
+ \om^{\,\,\,ac}_{\mu \,\cdot \cdot}\,
\om^{\,\,\,\,\,\, b}_{\nu c\, \cdot}
- \om^{\,\,\,ac}_{\nu \,\cdot \cdot}\,
\om^{\,\,\,\,\,\, b}_{\mu c\, \cdot}
\label{ncur}
\eeq
is a new notation, which becomes clear if we prove that there
is a direct relation with the Riemann tensor,
\beq
R_{\mu\nu ab}
&=&
R_{\mu\nu\rho\si}\,e^\rho_a\,e^\si_b\,.
\label{Rie}
\eeq
The proof consists of a direct substitution of Eqs.~(\ref{9}) with
the definition (\ref{spi}), in the expression (\ref{ncur}) and some
algebra which we leave as an exercise to the interested reader.

Using  (\ref{comm}) and  (\ref{Rie}), one can write a useful
covariant expression for the commutator,
\beq
 [\na_{\mu} ,\na_{\nu}] \psi
 &=&
-\,\frac14\,R_{\mu\nu\rho\si}\,\ga^\rho\ga^\si \,\psi\,.
\label{covcom}
\eeq

One of the applications of the previous expression is the possibility
of a ``doubling'' of the covariant Dirac equation (\ref{Dirac}).
Taking the product
\beq
\big( \ga^\mu \na_\mu - im \big)
\,\big( \ga^\nu \na_\nu + im \big)
&=&
\ga^\mu \na_\mu \,\ga^\nu \na_\nu + m^2\,,
\label{doubl}
\eeq
one can obtain
\beq
\ga^\mu \na_\mu \,\ga^\nu \na_\nu
&=&
\frac12\,\ga^\mu\ga^\nu
\big( \na_\mu \na_\nu + \na_\mu \na_\nu \big)
\,+\,
\frac12\,\ga^\mu\ga^\nu
\big( \na_\mu \na_\nu - \na_\mu \na_\nu \big)
\nonumber
\\
&=&
\frac12\,\big(  \ga^\mu\ga^\nu + \ga^\nu\ga^\mu\big)
\na_\mu \na_\nu
\,+\,
\frac12\,
\big( \na_\mu \na_\nu - \na_\mu \na_\nu \big)
\nonumber
\\
&=&
g^{\mu\nu}\,\na_\mu \na_\nu
\,+\,
\frac18\,R_{\mu\nu\rho\si}\,\ga^\rho\ga^\si \ga^\mu\ga^\nu
\,\,=\,\,
\Box \,-\, \frac14\,R\,.
\label{L-form}
\eeq
In the last step we used an identity
\beq
R_{\mu\nu\rho\si}\,\ga^\rho\ga^\si \ga^\mu\ga^\nu
&=&
- 2R\,,
\label{iden}
\eeq
which can be easily proved by using covariant version of relation
(\ref{gammas}) and the algebraic properties of the Riemann
tensor. We leave the verification of this identity as one more
exercise for the interested reader.

The two relations, i.e., (\ref{covcom}) and the
Lichnerowicz formula (\ref{L-form}) play important roles in
differential geometry and quantum field theory in curved space.

\section{Local conformal transformation}
\label{s5}

It is important to consider the conformal transformation of
the metric and fermion field. For the sake of generality, we
consider the theory in $n$ spacetime (or Euclidean space, as
there is no difference at this level). The action is a direct
generalization of (\ref{Dirac}),
\beq
S_f &=&
i \int d^nx \sqrt{-g} \,
{\bar \psi}\big( \ga^\mu \na_\mu - im \big) \psi\,.
\label{Dirac-n}
\eeq
The definition of the gamma matrices and covariant derivatives
do not change and the unique difference comes from the algebra
of the gamma matrices.  For the {\it global} conformal transformation
\beq
g_{\mu\nu} \rightarrow g_{\mu\nu}\,e^{2\la}\,,
\qquad
\psi  \rightarrow \psi\,e^{d_\psi\,\la}\,,
\qquad
\bar{\psi}  \rightarrow \bar{\psi}\,e^{d_\psi\,\la}\,,
\qquad
\la=const,
\label{glob-n}
\eeq
we easily find that the conformal weight of the spinor field that
provides the invariance of the \textit{massless} part of the action
(\ref{Dirac-n}), is $d_\psi = \frac{1-n}{2}$. In what follows, we
assume this value. For obvious reasons, the conformal transformation
for the conjugated fermion ${\bar \psi}$ has the same conformal
weight as in the case of $\psi$.

Using the global symmetry as a hint, consider the
{\it local} conformal transformation
\beq
g_{\mu\nu} \,=\, {\bar g}_{\mu\nu}\,e^{2\si}\,,
\qquad
\psi  \,=\, \psi_*\,e^{d_\psi\,\si}\,,
\qquad
\bar{\psi}  \,=\,  \bar{\psi}_*\,e^{d_\psi\,\si}\,.
\label{loc}
\eeq
In what follows we will omit space-time  arguments, but always
assume that $\,\si=\si(x)$. Furthermore, all metric-dependent
quantities with bars are constructed using the fiducial metric
$\,{\bar g}_{\mu\nu}$. Also, we use compact notation for the
partial derivative, e.g., $\,\si_{,\la}=\pa_\la\si$. Let us note that
many formulas related to local conformal transformations of
curvature tensors, their contractions, etc, can be found in
\cite{Stud}.

The transformation of the elements of the action (\ref{Dirac-n})
provides
\beq
g^{\mu\nu} \,=\, {\bar g}^{\mu\nu}\,e^{-2\si}\,,
\quad
\sqrt{- {\bar g}} = \sqrt{- g}  \,e^{n\si}\,,
\quad
e^\mu_a\,=\, {\bar e}^\mu_a\,e^{-\si}\,,
\quad
e_\mu^b\,=\, {\bar e}_\mu^b\,e^{\si}
\label{simps0}
\eeq
and, consequently,
\beq
\ga^\mu\,=\, {\bar \ga}^\mu\,e^{-\si}\,,
\quad
\ga_\mu\,=\, {\bar \ga}_\mu \,e^{\si}\,.
\label{simps1}
\eeq
One can use this set of formulas to check the relation
$d_\psi = \frac{1-n}{2}$. For the Christoffel symbols
and spinor connection, Eqs.~(\ref{6}) and (\ref{spi}) give
\beq
\Ga^{\la}_{\al \be}
&=&
\bar{\Ga}^{\la}_{\al\be} + \de^{\la}_{\al}\si_{,\be}
+ \si_{,\al}\de^{\la}_{\be} - \si_{, \tau}
\bar{g}^{\la\tau}\bar{g}_{\al\be}\,,
\nonumber
\\
\om^{\,\,\,ab}_{\mu\,\cdot\cdot}
&=&
\bar\om^{\,\,\,ab}_{\mu\,\cdot\cdot}
 \,+\,
 (\bar{e}^a_{\mu}\,\bar{e}^{\la b}
 - \bar{e}^b_{\mu}\,\bar{e}^{\la a})\,\si_{,\la}\,.
\label{simps2}
\eeq

For the contraction, we get
\beq
\label{21}
\nonumber
&&
\ga^{\mu}\na_{\mu}\psi
\, = \,
\bar{\ga}^{\mu} e^{-\si}
\Big\{
\pa_{\mu}\psi
- \frac{i}{4}\,\bar{\om}^{\,\,\,ab}_{\mu\,\cdot\cdot}\,\si_{ab}\psi
- \frac{i}{4}\,\big(\bar{e}^a_{\mu}\bar{e}^{\la b}
- \bar{e}^{b}_{\mu} \bar{e}^{\la a} \big) \si_{,\la}
\cdot \frac{i}{2}\,\big(\ga_a\ga_b - \ga_b\ga_a)\psi \Big\}
\\
&&	
\quad
= \,\bar{\ga}^{\mu}\bar{e}^{-\si}\Big\{
\bar{\na}_\mu \psi  + \frac{1}{4} (\bar{\ga}_{\mu}\bar{\ga}^{\la}
- \bar{\ga}^{\la}\bar{\ga}_{\mu})\,\si_{,\la}\psi \Big\}
\\
&&	
\quad
= \,
e^{-\si}\Big\{
\bar{\ga}^{\mu}\bar{\na}_\mu \psi
\,+\, \frac{n-1}2\,\bar{\ga}^{\la}\,\si_{,\la}\psi
\Big\}
\,=\,
e^{-\frac{n-3}{2} \si}\,\bar{\ga}^{\mu} \bar{\na}_\mu \psi_*\,.
\label{conf}
\eeq
Substituting (\ref{conf}),  (\ref{loc}), and (\ref{simps0}) in
(\ref{Dirac-n}), we arrive at the transformation law
\beq
i \int d^4x \sqrt{-g} \,
{\bar \psi}\big( \ga^\mu \na_\mu - im \big) \psi
&=&
i \int d^4x \sqrt{-{\bar g}} \,
{\bar \psi}_*\big( {\bar \ga}^\mu {\bar \na}_\mu
- im \,e^\si\big) \psi_*\,.
\label{Dirac-conf}
\eeq
Thus, the massless action is invariant under the local
conformal transformation.

One can restore local conformal symmetry in the general action
(\ref{Dirac}), by replacing the mass by a scalar field $\,\ph$, i.e.,
forming the Yukawa interaction term. The detailed consideration
of this issue is beyond the scope of the present contribution and
one has to consult, e.g., \cite{OUP}.

\section{Energy-momentum tensor for the Dirac field}	
\label{s6}

Consider the derivation of the \textit{dynamical}\footnote{An
alternative definition of the  \textit{canonical} energy-momentum
tensor is based on the Noether' s theorem.
After this canonical energy-momentum is making
symmetric we arrive at  its Belinfante improvement.  This symmetric
version and the dynamical definition described below,
give the same results in all particular flat-space cases, regardless the
general proof of their equivalence is not knows, to the best of the
author's knowledge. In curves space the dynamical definition
is preferable for obvious reasons. }
$\,T_{\mu\nu}\,$ for the Dirac field. The definition of this
tensor is
\beq
\label{24}
T^{\mu\nu} &=&
- \frac{2}{\sqrt{-g}}\, \frac{\de S_f}{\de g_{\mu\nu}}\,.
\eeq
There are several nontrivial details in using this definition. First
of all, action (\ref{Dirac}) is constructed from the tetrad and not
from the metric. On top of that, this action is not Hermitian. In
order to fix the last issue, one can reformulate the action in the
equivalent Hermitian form,
\beq
\label{25}
S_f
&=&
\frac{i}{2}\int d^{4}x \sqrt{-g}\,\big(
\bar{\psi} \ga^{\mu}\na_{\mu}\psi
- \na_{\mu}\bar{\psi}\,\ga^\mu\psi + 2im \big).
\eeq
The difference with the original expression (\ref{Dirac}) is the
integrals of total derivative term. This integral is irrelevant for
the variational derivatives, but the action (\ref{25}) gives a
consistent result in a more straightforward way and we shall
use this form of the action.

Concerning the first problem, we shall need the variation of the
tetrad corresponding to the variation of the metric
\beq
g_{\al\be} \,\rightarrow\, g^\prime_{\al\be} =
g_{\al\be} + h_{\al\be}.
\label{var}
\eeq
The solution of this problem is \cite{BuSh84,Wood} (see also
Chapter 9 of \cite{book})
\beq
&&
e_\mu^{\prime\,a}\,=\,e_\mu^{\,a}
\,-\,\frac12\,e_\nu^{\,a}h^\nu_{\,\,\mu}
\,+\,\frac38\,e_\nu^{\,a}h^\nu_{\,\,\la}h^\la_{\,\,\mu}
\,+\,\dots
\nn
\\
&&
e_b^{\prime\,\al}\,=\,e_b^{\,\,\al}
\,+\,\frac12\,e_b^{\,\be}h_\be^{\,\,\al}
\,-\,\frac18\,e_b^{\,\be}h_\be^{\,\,\la}h_\la^{\,\,\al}
\,+\,\dots
\label{vartet}
\eeq
The first variations of other relevant quantities has the form
\beq
\nonumber
\de \sqrt{- g} &=& \frac{1}{2}\sqrt{- g} h\,,
\quad
\de g^{\mu\nu} = - h^{\mu\nu}\,,
\quad
\de e^{c}_\mu = \frac12\,h^\nu_\mu e^c_\nu\,,
\quad
\de e^\rho_b = - \frac12\,h^{\rho}_{\la} e^\la_{b}\,,
\\
\label{26}
\de \Ga^{\la}_{\al \be}
&=&
\frac12\,
\big(
\na_{\al}h^\la_\be + \na_\be h^\la_\al - \na^\la
h_{\al \be}\big)\,,
\quad
\de \ga^\mu = - \frac12\,h^\mu_\nu\,\ga^\nu\,.
\eeq
Here all indices are raised and lowered with the
background metrics $g^{\mu\nu}$ and $g_{\mu\nu}$
and $h=g^{\mu\nu}h_{\mu\nu}$.
Furthermore, direct calculation using (\ref{spi}) and
(\ref{26}) yields
\beq
\de \om^{\,\,\,ab}_{\mu \,\cdot \cdot}
\,=\,
- \,\de \om^{\,\,\,ba}_{\mu \,\cdot \cdot}
\,=\,\frac{1}{2}\,\de (e^{b}_\tau e^{\la a}
\Ga^\tau_{\la \mu} - e^{\la a}\pa_{\mu}e^{b}_{\la})
\,=\,
\frac12\,
\big(e^{a\tau}e^{b\la} - e^{b\tau}e^{a\la}\,)
\na_{\la}h_{\mu \tau} \,.
\label{27}
\eeq
One more useful relation is
\beq
\bar{\psi}\ga^{\mu}\na_{\mu}\psi
- \na_{\mu}\bar{\psi}\ga^{\mu}\psi
\,=\, \bar{\psi}\ga^{\mu}\pa_{\mu}\psi - \pa_{\mu}\bar{\psi}\ga^{\mu}\psi
- \frac{i}{4}\,\om^{\,\,\,ab}_{\mu\, \cdot \cdot}
\,\bar{\psi}\big(
\ga^{\mu}\si_{ab} + \si_{ab}\ga^{\mu}\big)\psi\,.
\eeq
Consider the term depending on the variation of
$\om^{\,\,\,ab}_{\mu\, \cdot \cdot}$, defined by (\ref{27}),
\beq
\de_{\om}S_f
&=&
- \frac{i}{4}\,\int d^4x\sqrt{-g}
\,\bar{\psi}\big(
\ga^{\mu}\si_{ab} + \si_{ab}\ga^{\mu}\big)\psi
\,\de \om^{\,\,\,ab}_{\mu\, \cdot \cdot}
\nonumber
\\
&=&
 - \frac{i}{4} \int d^{4}x \sqrt{-g}\,
 \frac12\,\,\big(e^{a\tau}e^{b\la} - e^{b\tau}e^{a\la}\big)
 \,\big( \na_\la h_{\mu \tau}\big)\,e^{\mu c}\,
 \bar{\psi}(\ga_{c}\si_{ab} + \si_{ab}\ga_{c})\psi\,.
\eeq
After integration by parts, we get
\beq
\de_{\om}S_f
&=&
\frac{i}{8} \int d^4x \sqrt{-g} \,
h_{\mu \tau} e^{\mu c}
\big(e^{\tau a}e^{\la b} - e^{\tau b}e^{\la a}\big)\,
\na_{\la}\big[
\bar{\psi }(\ga_{c}\si_{ab} + \si_{ab}\ga_{c})\psi\big]\,.
\eeq
It is important that the term in the brackets $[...]$ is
antisymmetric in $\,ab$. Hence one can simplify the last
expression, using the expression for $\si_{ab}$,
\beq
\de_{\om}S_f
\,=\,
-\,\frac{1}{8} \int d^{4}x \sqrt{-g} \,h_{\mu\nu} \,
e^{\mu c} e^{\nu a} e^{\la b} \na_{\la} \big[
\bar{\psi}(\ga_c\ga_a\ga_b - \ga_c\ga_b\ga_a
+ \ga_a\ga_b\ga_c - \ga_b\ga_a\ga_c)\psi\big]\,.
\nonumber
\eeq
Replacing $\,\ga_{a}\ga_{b} + \ga_{b}\ga_{a} = 2 \eta _{ab}$,
we arrive at
\beq
\label{28}
\de_{\om}S_f
\,=\,
-\,\frac{1}{8}  \int d^4 x \sqrt{-g} \, h_{\mu\nu} \na_\la
\big[ \bar{\psi}
(g^{\la\mu}\ga^\nu - g^{\la\nu}\ga^\mu)\psi\big]
 \,=\, 0\,.
\eeq
Thus, the
$\,\de\om^{\,\,\,ab}_{\mu \,\cdot \cdot}\,$ is irrelevant as
its contribution vanish. Hence we arrive at
\beq
\de S_f
&=&
\frac{i}{2}\int d^{4}x \sqrt{-g}
\Big\{ \frac12\,\big( h g^{\mu \nu}  - h^{\mu \nu}\big)
\big(\bar{\psi} \ga_\nu \na_\mu \psi
- \na_\mu \bar{\psi} \ga_\nu \psi\big)
- ihm\bar{\psi}\psi \Big\}
\nonumber
\\
&&
=\,\,
 \int d^4x \sqrt{-g}\,
 h_{\al \be}\,
 \Big\{ \frac{i}{4} g^{\al \be}
\big(
\bar{\psi}\ga^\la \na_\la \psi - \na_\la\bar{\psi} \ga^\la\psi
\big)
- \frac12\,g^{\al \be}m \bar{\psi}\psi
\nonumber
\\
&&
-\,\,\frac{i}{4}\,\big(\bar{\psi}\ga^\al\na^\be\psi -
 \na^\al \bar{\psi}\ga^\be \psi\big)\Big\}\,.
\label{delS}
\eeq
The expression (\ref{delS}) is interesting by itself, as it describes
an interaction of massive Dirac fermion with a weak gravitational
field. In particular, it can be used for elaborating the
nonrelativistic approximation to the fermions, or for deriving
the equations of motion for the spinning charged classical particle
in a weak relativistic field \cite{Goncalves-2007,Quash-2015}
(also the forthcoming work \cite{ePaePa} with more references).

On the other hand, using the definition (\ref{24}), the
energy-momentum tensor of the fermion has the form
\beq
T_{\mu\nu}
\,=\,
\frac{i}{2}\,\big[\bar{\psi}\ga_{(\mu} \na_{\nu)}\psi
- \na_{(\mu}\bar{\psi}\ga_{\nu)}\psi\big]
- \frac{i}{2} \,g_{\mu\nu}
\big[
\bar{\psi}\ga^\la \na_\la \psi - \na_\la\bar{\psi} \ga^\la\psi
\big]
\,+\,m \bar{\psi}\psi\,g_{\mu\nu}.
\label{Tmn}
\eeq

Let us take a trace of the tensor (\ref{Tmn}). We get
\beq
T_\mu^\mu
&=&
T_{\mu\nu} g^{\mu\nu}
\,=\,
- \frac{3i}{2}\,\big(
\bar{\psi}\ga^\la \na_\la \psi - \na_\la\bar{\psi} \ga^\la\psi
\big)
\,+\,4m \bar{\psi}\psi\,.
\label{traceTmn}
\eeq
Using the equations of motion
\beq
\ga^\mu \na_\mu\psi = - im \psi
\quad
\mbox{and}
\quad
\na_\mu \bar{\psi}\ga^\mu  \,=\,  im \bar{\psi},
\label{eqs}
\eeq
the on shell trace is
\beq
T_\mu^\mu\Big|_{on\,\, shell}
&=&
m \bar{\psi}\psi\,.
\label{trace on-shell}
\eeq
For a massless theory this expression vanishes, in accordance
with our previous finding that the massless fermion theory is
invariant under local conformal transformation.

There is an alternative way of deriving the trace (\ref{traceTmn}).
Let us remember that the definition (\ref{24}) is valid for any kind
of a matter field $S_m$. Take a variational derivative of the action
with respect to the conformal factor $\si$ and using (\ref{loc}),
we get
\beq
\frac{\de' S_m}{\de \si}
\,=\,
\frac{\de S_m}{\de g_{\mu\nu}}\,\frac{\de g_{\mu\nu}}{\de \si}
\,=\,
\frac{\de S_m}{\de g_{\mu\nu}}\,2\,
\bar{g}_{\mu\nu}\,e^{2\si}
\,=\,
\frac{\de S_m}{\de g_{\mu\nu}}\,2\,
g_{\mu\nu}
\,=\,
- \,\sqrt{-g} \, T^{\mu\nu}\,g_{\mu\nu}.
\label{Noethercon}
\eeq
Here the prime over variation in the first expression means that
we take into account only the dependence of the metric and not
of the matter fields. Formula (\ref{Noethercon}) is valid, in
particular, for the Dirac fermions in curved space, and can be used
for getting $T^\mu_\mu$.

Adding variations with respect to the matter fields, we get
\beq
\frac{1}{\sqrt{-g}}\,\frac{\de S_m}{\de \si}
\,=\,
\frac{1}{\sqrt{-g}}\Big(
2 g_{\mu\nu}\,\frac{\de S_m}{\de g_{\mu\nu}}\,
\,+\, \sum\limits_k d_k \,\Phi_k
\frac{\de S_m}{\de \Phi_k}\Big),
\label{gencon}
\eeq
where $d_k$ is the conformal weight of the field $\Phi_k$
(e.g., $d_k=-1$ for scalars and $d_k=-3/2$ for spinors in $n=4$).
Obviously, 
(\ref{gencon}) is equivalent to
(\ref{Noethercon}) on shell, when  $\de S_m/\de \Phi_k=0$.
Thus, the expression for the trace (\ref{trace on-shell}) can be
obtained, even for the massive nonconformal case, by the following
sequence of steps: \ \textit{i)}
Rewriting the action (\ref{Dirac}) using the parametrization
(\ref{loc});  \ \textit{ii)} Taking the variational derivative
with respect to $\si$;  \ \textit{iii)} Replacing $\si \to 0$ and
$\bar{g}_{\mu\nu} \to g_{\mu\nu}$ and (off shell) the
same for the spinor field.

The  considerations presented above and its result (\ref{gencon})
are valid for all matter fields. In the conformal case, the
\textit{r.h.s.} is zero, which establishes the relation between local
conformal symmetry (i.e., independence on $\si$) and the vanishing
on the mass shell expression for the trace of the energy-momentum
tensor.

\section{Derivation of $T_{\mu\nu}$ on the cosmological
background}
\label{s7}

In order to have an illustration for the results presented above,
let us calculate the energy-momentum tensor for a free spinor
on a cosmological background. It is assumed that not only metric,
but also the fermion field depends only on time and not on the
space coordinates.
For the sake of simplicity we choose a conformally-flat metric
\beq
g_{\mu\nu} = a^2(\eta)\bar{g}_{\mu\nu}
&=& e^{2\si(\eta)}\bar{g}_{\mu\nu}\,,
\eeq
with $\,\bar{g}_{\mu\nu} = \eta_{\mu\nu}$. We
also use the conformal time variable $\eta$, related to
physical time $\,t\,$ as $\,d\eta = a(t)dt$. The derivative
with respect to $\eta$ will be denoted by prime. It is
well-known that the non-zero components of affine
connection are
\beq
\Ga^0_{00} = \frac{a^\prime}{a}
\,,\qquad
\Ga^0_{ik} = \frac{a^\prime}{a}\,\de_{ik}
\,,\qquad
\Ga^i_{0k} = \frac{a^\prime}{a}\,\de^i_k\,.
\label{afficos}
\eeq
Here and in what follows we shall use the notations
$\,i,j,k=1,2,3\,$ and $\,a,b,c=0,1,2,3$, while both sets of
Latin indices correspond to the flat fiducial metric. The
tetrad components can be easily derived and the
curved-space gamma-matrices are given by
\beq
\ga^0 = \frac{1}{a}\,\Ga^0
\,,\qquad
\ga_0 = a\,\Ga^0
\,,\qquad
\ga^k = \frac{1}{a}\,\Ga^k
\,,\qquad
\ga_k = a\,\Ga_k\,,
\label{gam}
\eeq
where the useful notations $\Ga^a = \ga^a$ were introduced
for the flat-space gamma's.

In order to derive the components of $\,T_{\mu\nu}\,$
one has to take care  about the spinor connection first.
Direct calculation gives (other components are equal to zero)
\beq
\nonumber
\om^{\,\,\,k 0}_{0\,\cdot\cdot}
&=&
\Ga^{k}_{0 0} - e^{-\si}\eta^{\la 0}\pa_{0}
\big(e^{\si}\de^{k}_{\la}\big)
= \Ga^{k}_{00} = 0\,,
\\
\nonumber
\om^{\,\,\,k i}_{0\,\cdot\cdot}
&=&
\Ga^{\tau}_{\la 0}e^{k}_{\tau}e^{\la i}
- e^{\la i} \pa_0 e^k_\la
\,=\,
- \frac{a'}{a} \eta^{k i} \,=\, \frac{a'}{a}\,\de^{k i}\,,
\\
\nonumber	
\om^{\,\,\,i k}_{j \,\cdot\cdot}
&=&
\Ga^{\tau}_{\la j}e^{i}_{\tau}e^{\la k} - e^{k \la} \pa_{j}e^{i}_{\la} = 0\,,
\\
\om^{\,\,\,o i}_{k\,\cdot\cdot}
&=&
\Ga^{\tau}_{\la k}e^{o}_{\tau}e^{\la i} - e^{\la i}\pa_{k}e^{0}_{\la} \,=\,
\Ga^{0}_{jk}\eta^{ij} = \frac{a'}{a}\eta_{jk}\eta^{ij}
= - \frac{a}{a}\de^{i}_{k}\,.
\label{wcosmo}
\eeq
Thus, we arrive at
\beq
\na_{0}\psi
&=&
\psi' - \frac{i}{4}\om_{0}^{kj}\si_{kj} \psi
= \psi' + \frac{i}{4}\,\frac{a'}{a}\,\eta^{kj}\si_{kj} = \psi'\,.
\nonumber
\eeq
Similarly, $\,\na_{0}\bar{\psi} = \bar{\psi}'$.
Furthermore, it is easy to obtain
\beq
\na_i\psi
&=&
\pa_i \psi - \frac{i}{2}\om_{i \,\cdot \cdot}^{\,\,\,ok} \si_{ok}\psi
= \frac{i}{2}\,\frac{a'}{a}\,\frac{i}{2}\,\,
2\Ga_{0}\Ga_{k}\psi\de
\,=\, -\frac{a'}{2a}\Ga_{o}\Ga_{i}\,,
\nonumber
\\
\na_i\bar{\psi}
&=&
\pa_i\bar{\psi}
+ \frac{i}{2}\,\om_{i\,\cdot\cdot}^{\,\,\,0 k} \bar{\psi} \,\si_{ok}
\,=\,
\frac12\,\de_{i}^{k}\,\frac{a'}{a}\,\bar{\psi} \Ga_{0}\Ga_{k}
\,=\, +\,\frac{a'}{2a}\,\bar{\psi}\Ga_{o}\Ga_{i}\,.
\nonumber
\eeq

Now we are in a position to calculate the components of  $T_{\al\be}$.
According to Eq. (\ref{Tmn}),
\beq
T_{00}
&=&
- \frac{i}{2}\,g_{00} \big(\bar{\psi}\ga^{0}\na_{0}\psi
+ \bar{\psi}\ga^{k}\na_{k}\psi
-  \na_{0}\bar{\psi}\ga^{0}\psi
-  \na_{k}\bar{\psi}\ga^{k}\psi\big)
+  g_{00}\bar{\psi}\psi m
\nonumber
\\
&+&
\frac{i}{2}\,\big( \bar{\psi}\ga_{0}\na_{0}\psi - \na_{0}\bar{\psi}\ga_{0}\psi\big)
\,=\, a^{2}m \bar{\psi} \psi\,.
\label{33}
\eeq
In a similar way, taking into account $g_{0k}=0$, we get
\beq
T_{0k}
&=&
\frac{i}{4}\,
\big(\bar{\psi}\ga_{0}\na_{k}\psi
+ \bar{\psi} \ga_{k}\na _{0}\psi
- \na_{0}\bar{\psi}\ga_{k}\psi
- \na_{k}\bar{\psi}\ga_{0}\psi\big)
\nonumber
\\
&=&
\frac{ia}{4}\,\big(
\bar{\psi}\Ga_{k}\psi' - \bar{\psi}'\Ga_{k}\psi\big)\,.
\label{34}
\eeq
Because of the homogeneity and isotropy of space, in the cosmological
setting this must be zero. Let us show that this is the case.

On the mass shell, we have
\beq
\ga^{0}\na_{0}\psi + \ga^{\mu}\na_{k}\psi + im\psi
&=&
\frac{1}{a}\,\Ga^0
\psi' - \frac{a'}{2a^2}\,\Ga^k\Ga_0\Ga_k\psi
+ im\psi = 0\,.
\eeq
Since $\,\Ga^{k}\Ga_{0}\Ga_{k} = - 3\Ga_{0}, $ we get
\beq
\frac{1}{a}\Ga^0
\big(
\psi' + \frac{3}{2}\frac{a'}{a}\psi\big) + im\psi \,=\, 0\,.
\label{37a}
\eeq
Similarly, starting from
$\,\na_{\mu}\bar{\psi}\ga^{\mu} - im\psi = 0$,
one can arrive at 	
\beq
\frac{1}{a}\,\Big(
{\bar \psi}'
+ \frac{3a'}{2a} \bar{\psi}\Big)\Ga_0
\,-\, im\bar{\psi} = 0 \,.
\label{37}
\eeq
Then, (\ref{34}) becomes the following on shell expression:
\beq
\label{36}
T_{0k}\Big|_{on\,\,shell}
&=&
- \frac{ia}{4}\,ma \big(
\bar{\psi}\Ga_{k}\Ga_{0}\psi
+ \bar{\psi}\Ga_{0}\Ga_{k}\psi\big) \,=\, 0\,,
\eeq
exactly as required to guarantee the vanishing non-diagonal
element (\ref{34}).

The last part is to derive the space components,
\beq
\nonumber
T_{ki}
&=&
- ia^2\eta_{ik} \big(
\bar{\psi}\ga^{0}\na_{0}\psi
+ \bar{\psi}\ga^{j}\na_{j}\psi - \na_{0}\bar{\psi}\ga^{0}\psi
- \na_j\bar{\psi}\ga^{j}\psi + 2im\bar{\psi}\psi\big)
+ \frac{i}{2}\big[\bar{\psi}\ga_{(k}\na_{i)}\psi\big]
\\
\nonumber
&=&
- \frac{ia}{2}\eta_{ik}
\big(
\bar{\psi}\Ga^{0}\psi' - \bar{\psi}'\Ga^{0}\psi - \frac{a'}{2a}
\bar{\psi}\Ga^{j}\Ga_{0}\Ga_{j}\psi - \frac{a'}{2a}\bar{\psi}
\Ga_{0}\Ga_{j}\Ga^{j}\psi\big)
\\
\nonumber
&+&
a^{2}\eta_{ik}\,m\bar{\psi}\psi
- \frac{i a'}{8}\,\big[\bar{\psi}\Ga_{(k}\Ga_{|0|}\Ga_{i)}\psi
+ \bar{\psi}\Ga_{0}\Ga_{(k}\Ga_{i)}\psi\big]
\\
\nonumber
&=&
-\frac{ia}{2}\,\eta_{ik}\,
\big(\bar{\psi}\Ga_{0}\psi' - \bar{\psi}'\Ga_{0}\psi\big)
+ ma^2\,\eta_{ik}\bar{\psi}\psi
- \frac{ia'}{a}\,\bar{\psi}
\big(\Ga_{0}\Ga_{(k}\Ga_{i)} - \Ga_{0}\Ga_{(k}\Ga_{i)}\big)
\psi
\\
&=&
\frac{ia}{2}\,\eta_{ik}\,\big(
\bar{\psi}'\Ga_{0}\psi - \bar{\psi}\Ga_{0}\psi'\big)
\,+\, ma^{2}\eta_{ik}\bar{\psi}\psi\,.
\label{38}
\eeq
	
On shell we get, according to (\ref{37a}) and (\ref{37}),
\beq
T_{ik}\Big|_{on \,\,shell}
&=&
\eta_{ik}\,\Big[ ma^{2} \bar{\psi}\psi
+ \frac{a^2}{2}\,m\,
\big(\bar{\psi}\Ga_{0}^{2}\psi + \bar{\psi}\Ga_{0}^{2}\psi
\big) \Big] \,=\,0\,.
\label{39}
\eeq
Here we used the $\Ga_{0}^{2}=1$ property of the gamma-matrix.

For a massive case,
we meet $ T_{11} = T_{22} = T_{33} = 0$,
while $\,\rho = T_{00} \neq 0$. This is a dust-type equation of
state for the matter.

The physical interpretation of the results for diagonal component
(\ref{33}) and (\ref{39}) is the following. First of all, for a massless
case, $\,T_{\mu\nu}\big|_{on\,\,shell} = 0\,$ for the FRW metric. This
is a natural output, since we permitted the fermion field to depend
only on conformal time and {\it not} on the space coordinates.
Therefore, we exclude the field configurations such as plane (and
other, of course) waves. In the massive case, there is still a dust-like
configuration and in the massless version no solutions are possible.
Of course, if we permit the space-dependence, the equation of state
will be the one of radiation,
$\,T_\mu^\al = \diag \big(\rho_r\,,\,-\frac13 \de_i^j \rho_r\big)$.

\section{Conclusions}
\label{s8}

We presented in details the derivation of spinor connection,
covariant derivative of Dirac fermion, and energy-momentum tensor
for the fermion. Also, the conformal properties of fermions in
curved space were discussed.
Let us note that the same constructions can be applied to more
general backgrounds, for instance for the non-Riemannian space
which has not only metric, but also metric-independent torsion.
The result can be found, e.g., in \cite{book} or in the review papers
\cite{Hehl-76,torsi}. In a more general case, when the metricity
condition is not  satisfied, the construction of covariant derivative
of the fermions meets difficulties and the output of this procedure
looks unclear, at the moment. In principle, the scheme described
here should work is this and other cases, when the geometry is
enriched by additional metric-independent fields.

The last observation is that the importance of the covariant
formulation of spinors in curved space goes beyond the quantum
field theory. Regardless spin is certainly a property of quantum
theory, the fermions in curved space are important for the models
of spinning particles and the last can be used as basic models for
the macroscopic gravitationally interacting compact magnetic bodies.
This is an important issue in view of existing experimental efforts
(see, e.g., \cite{South} and references therein) towards the precise
measurements of the  accelerations of such particles. At the moment,
the existing theoretical literature on this issue is restricted to
elementary spinning particles
\cite{Obukhov-2000,Silenko-2006,Gosselin-2010,Obukhov-2013}
(see also the planned continuation of the present work in
\cite{ePaePa}), but it would be certainly interesting to extend the
existing understanding to the macroscopic bodies.

\section*{Appendix. Gamma matrices}

The three Pauli matrices are defined as
\beq
\sigma_1=\left(\begin{array}{cc}
0 & 1 \\
1 & 0
\end{array}
\right),
\qquad
\sigma_2=\left(\begin{array}{cc}
0 & -i  \\
i &  0
\end{array}
\right),
\qquad
\sigma_3=\left(\begin{array}{cc}
1&0\\
0&-1
\end{array}
\right).
\qquad
\eeq

In the standard representation, the gamma (Dirac) matrices
are defined as
\beq
\gamma^a \,=\,
\left(\begin{array}{ccc}
0            & \sigma^a \\
- \sigma^a &  0 \end{array}
\right),
\label{gamma}
\eeq
satisfying the Clifford algebra
\begin{eqnarray}
&&
\ga^a \ga^b + \ga^b \ga^a
\,=\,2\eta^{ab} I,
\label{Clifford}
\end{eqnarray}
where $I$ is the four-dimensional unit matrix.

It is important that neither Pauli matrices nor Dirac matrices do not
form a vector with respect to the Lorentz transformation and
are just constant matrices which do not change under Lorentz
transformation. This important feature is compatible with
(\ref{Clifford}) and with the Lorentz covariance of the Dirac
equation \cite{BjorkenDrell-1,OUP}. It also holds in other
representations of $\ga^a$. It is interesting that all this is true
only in the Cartesian coordinates in the space section. To write
the gamma matrices, e.g., in spherical coordinates one has to
use the tetrads, with the corresponding change in (\ref{Clifford}).

\section*{Acknowledgements}

 I am grateful to Prof. Miguel Gustavo de Campos Batista for an
 assistance in typing  the first draft of this work back in 2016.
The contribution of Samuel William de Paulo Oliveira and
Guilherme Yoshi Oyadomari, who found a few important
 misprints in the original version, is greatly appreciated.
The author is partially supported by Conselho Nacional de
Desenvolvimento Cient\'{i}fico e Tecnol\'{o}gico - CNPq (Brazil),
the grant 303635/2018-5 and by Funda\c{c}\~{a}o de Amparo \`a
Pesquisa de Minas Gerais - FAPEMIG, the project PPM-00604-18;
and by the Ministry of Education of the Russian Federation, under
the project No. FEWF-2020-0003.


\end{document}